\documentstyle[aas2pp4,epsf]{article}

\def\bea{\begin{eqnarray}}
\def\eea{\end{eqnarray}}
\def\nn{\nonumber \\[1mm]}

\begin{document}

\title{Radiation of Angular Momentum by Neutrinos from Merged Binary 
	Neutron Stars}
\author{Thomas W.~Baumgarte and 
	Stuart L.~Shapiro\altaffilmark{1}}
\affil{Department of Physics, University of Illinois at
        Urbana-Champaign, Urbana, Il~61801}
\altaffiltext{1}{Department of Astronomy and
        National Center for Supercomputing Applications, 
        University of Illinois at Urbana-Champaign, Urbana, Il~61801}

\begin{abstract}
We study neutrino emission from the remnant of an inspiraling binary
neutron star following coalescence. The mass of the merged remnant is
likely to exceed the stability limit of a cold, rotating neutron star.
However, the angular momentum of the remnant may also approach or even
exceed the Kerr limit, $J/M^2 = 1$, so that total collapse may not
be possible unless some angular momentum is dissipated.
We find that neutrino emission is very inefficient in decreasing the
angular momentum of these merged objects and may even lead to a small
increase in $J/M^2$.  We illustrate these findings with a
post-Newtonian, ellipsoidal model calculation.  Simple arguments suggest
that the remnant may form a bar mode instability on a timescale
similar to or shorter than the neutrino emission timescale, in which case
the evolution of the remnant will be dominated by the emission of
gravitational waves.
\end{abstract}

\section{INTRODUCTION}

Considerable theoretical effort is currently directed toward
understanding the coalescence of binary neutron stars. The interest
stems partly from the promise such coalescence holds for generating
gravitational waves that can be detected by laser interferometers now
under construction, like LIGO, VIRGO and GEO. Additional interest is
triggered by the idea that neutrinos generated during the merger and
coalescence might be the source of gamma rays bursts (see, e.g.,
Paczy\'nski, 1986; Eichler {\em et.~al.}, 1989; Janka \& Ruffert,
1996).  Most of the theoretical work has, to date, been performed in a
Newtonian framework (see, e.g., Rasio \& Shapiro, 1994; Ruffert, Janka
\& Sch\"afer 1996; Zhuge, Centrella \& McMillan, 1996). Only a few
hydrodynamical calculations have taken into account post-Newtonian
corrections (Oohara \& Nakamura, 1995; Shibata, Oohara \& Nakamura,
1997), while results from calculations in full general relativity are
very preliminary (Oohara
\& Nakamura, 1995; Wilson \& Mathews, 1995; Wilson, Mathews \& 
Marronetti, 1996).

There are a number of issues, however, that cannot be addressed even
qualitatively in Newtonian gravitation. Such issues include the
collapse of the merged remnant to a black hole. Such a collapse is
quite likely because the mass of the remnant will probably exceed the
maximum allowed mass for a single neutron star. However, the final fate of the
collapse is completely unknown if the remnant's angular momentum
parameter $J/M^2$ exceeds unity. In this case, the remnant cannot form
a Kerr black hole without first losing some of its angular
momentum. In principle, angular momentum could be radiated away by the
emission of gravitational waves, neutrinos, and/or electromagnetic
(e.g. magnetic dipole) radiation.

During the coalescence of the two neutron stars some fraction of the
kinetic energy will be converted into thermal energy. Therefore the
remnant is likely to be hot and emit thermal neutrinos.  During the
rapid plunge of the binary once it reaches the innermost stable
circular orbit (ISCO), the angular momentum of the system will be
conserved approximately, so the remnant will be rotating very
rapidly. In their fully relativistic treatment of corotating,
polytropic binaries in quasiequilibrium circular orbit, Baumgarte {\em
et.~al.} (1997a, 1997b) found that, for a polytropic equation of state
with polytropic index $n=1$, the total $J/M^2$ of the binary system at
the ISCO exceeds unity for all binaries in which each member has a
rest mass $\lesssim 0.94 M_0^{\rm max}$.  Here $M_0^{\rm max}$ is
the maximum allowed rest mass of a nonrotating, isolated, cold neutron
star, and $M_0$, $M$ and $J$ denote the total rest mass, energy and
angular momentum of the binary system.  For more realistic,
nonsynchronous binaries, $J/M^2$ is likely to be somewhat smaller, but
may still be appreciable.  For many of these binaries the rest mass of
the merged remnant also exceeds $M_0^{\rm max}$. Immediately following
merger, these objects may be temporarily stabilized by thermal
pressure and angular momentum. Only after some of the thermal energy
and angular momentum have been radiated away can the star later
collapse to a black hole.  The ultimate fate of such an object can
only be established by a three-dimensional, relativistic hydrodynamic
numerical simulation which employs a hot nuclear equation or state and
full neutrino radiation transport. While the development of such a
code is underway by several groups, it is still very far from
complete.

Our goal in this paper is to isolate the role of the neutrino emission
in the early evolution following merger and to estimate its effect on
the angular momentum $J$ and the angular momentum parameter
$J/M^2$. To do so, we consider a simplified scenario for the
post-merger neutron star remnant and imagine that the remnant evolves
to an axisymmetric, quasi-equilibrium state on a dynamical (orbital)
timescale immediately following merger. Since the merger follows a
rapid plunge from the ISCO, the merged remnant forms with the same
total mass $M$, the rest mass $M_0$ and the angular momentum $J$ of
the progenitor binary at the ISCO (here we neglect the small loss of
mass and angular momentum via gravitational waves and gas ejection
during the dynamical plunge).  Accordingly, $J/M^2$ will typically be
near unity. We assume that the merged configuration quickly settles
into uniformly rotating, dynamically stable quasiequilibrium state and
focus on its quasi-static evolution.  We show that the emission of
neutrinos is very inefficient in carrying off $J$. Depending on the
stiffness of the equation of state, this emission may lead to either
an increase or a decrease of $J/M^2$. Also, we argue that these
objects typically develop a bar instability on a timescale comparable
or shorter than the neutrino emission timescale. We therefore conclude
that while neutrino emission may be important for the cooling of the
remnant, it is negligible for the emission of angular momentum.

The paper is organized as follows. In Section~2 we estimate the rate
of angular momentum loss by neutrino emission. In Section~3 we
incorporate our estimate in a post-Newtonian, ellipsoidal model
calculation, that follows the post-merger evolution of the
remnant. The equations and formalism for this approximation are
presented in Appendix~A. In Section~4 we compare the relevant
timescales for angular momentum loss by competing processes, and we
summarize our conclusions in Section~5.

\section{ANGULAR MOMENTUM LOSS BY \break NEUTRINO EMISSION}

Consider a neutrino $\nu$ with four-momentum $p^{\alpha}$ emitted from
the surface of a uniformly rotating axisymmetric equilibrium star into
an arbitrary direction. A local observer with four velocity
$u^{\alpha}$ comoving with the surface will measure the neutrino's
energy to be
\begin{equation}
\delta W^{(e)} = - u^{\alpha} p_{\alpha}
	= - u^t p_t - u^{\varphi} p_{\varphi}.
\end{equation}
Here and throughout the paper we adopt geometrized units, $c \equiv 1
\equiv G$.  Since the spacetime is stationary and axisymmetric, the
components $p_t$ and $p_{\varphi}$ are conserved along the neutrino's
path.  Therefore, a distant observer at rest with respect to the
star's center of mass will measure the neutrino's energy to be
\begin{equation}
\delta W^{(r)} = - p_{t} = - dM,
\end{equation}
A distant observer will identify the neutrino's
angular momentum to be
\begin{equation}
dJ = - p_{\varphi}.
\end{equation}
Combining the last three equations yields the familiar result (e.g., 
Thorne, 1971)
\begin{equation} \label{fl0}
dM = \Omega dJ - \frac{1}{u^t} \delta W^{(e)},
\end{equation}
where $\Omega = u^{\varphi}/u^t$ is the angular velocity of the
rotation.  The luminosity as measured by a local, comoving observer
can be defined as
\begin{equation} \label{defL}
L = \frac{\delta W^{(e)}}{d \tau},
\end{equation}
where $\tau$ is the observer's proper time. Equation~(\ref{fl0}) can then
be rewritten
\begin{equation} \label{fl}
\dot M = \Omega \dot J - \frac{1}{u^t} L,
\end{equation}
where the dot denotes a derivative with respect to $\tau$.

Equation~(\ref{fl}) is fully relativistic. The relation between $L$
and $\dot J$ depends on the characteristics of the emission of the
neutrinos from the star's surface, as well as the relativistic
structure of the star. Establishing this relationship therefore
requires detailed numerical models of radiating, relativistic,
rotating stars.  Instead, we adopt a simple, first-order Newtonian
description to find the approximate relationship given by
equation~(\ref{dotJ2}) below.  We assume that the neutrinos are
emitted isotropically in the rest frame of a local stationary observer
comoving with the surface. An observer in a nonrotating static frame
will therefore find that, on average, each neutrino carries off some
angular momentum. This angular momentum can be found by Lorentz
transforming from the stationary frame into the static frame. To
lowest order in $v$ we find that, on average, each neutrino carries
off a linear momentum $p_{\hat \varphi} = v E_{\nu} = \Omega\,
\varpi E_{\nu}$, where $\Omega$ is the angular velocity of the star,
$\varpi$ the distance of the surface element from the axis of
rotation, and $E_{\nu}$ is the energy of the neutrino in the comoving
frame (equation~(1)).  The corresponding angular momentum is
\begin{equation}
p_{\varphi} = \varpi^2\,\Omega E_{\nu}.
\end{equation}
The total rate of loss of angular momentum is therefore (see also
Kazanas, 1977)
\begin{equation} \label{dotJ1}
\dot J = \,- <\varpi^2> \Omega L,
\end{equation}
where $L$ is the neutrino luminosity and $<\varpi^2>$ denotes the
average of $\varpi^2$ over the surface of the star. We can get a 
reasonable estimate for this average by adopting a rotating,
compressible ellipsoid model for the star (see Lai, Rasio
\& Shapiro, 1993, hereafter LRS). If we denote the principal axes by
$a_1$, $a_2$ and $a_3$ (where $a_1 = a_2$ is measured in the
equatorial plane and $a_3$ is along the rotation axis), we find
\begin{equation}
<\varpi^2>\, = \frac{2}{3} a_1^2 \,f(e),
\end{equation}
where 
\bea
f(e) & \equiv & \frac{6 \pi}{a_1^2 {\cal A}}\, 
	\int^{a_1}_{0} \varpi^3 \left( 		
	\frac{a_1^2 - e^2 \varpi^2}{a_1^2 - \varpi^2}
	\right)^{1/2} d\varpi  \nn
	& = & 1 - \frac{1}{15} e^2 + O(e^4).
\eea
Here 
\begin{equation}
e^2 \equiv 1 - (a_3/a_1)^2
\end{equation}
is the eccentricity, and
\begin{equation}
{\cal A} = 4 \pi \int^{a_1}_{0} \varpi \left( 
		\frac{a_1^2 - e^2 \varpi^2}{a_1^2 - \varpi^2}
			\right)^{1/2} d\varpi
\end{equation}
is the surface area of the ellipsoid. Note that $f(e) \sim 1$ even for
large eccentricities.  It is convenient to rewrite
equation~(\ref{dotJ1}) in terms of the angular momentum $J$ and the
moment of inertia $I$. The latter is $I = 2/5\,\kappa_n M a_1^2$,
where $\kappa_n$ is a dimensionless structure constant of order unity
which depends on the star's density profile. For a polytropic equation
of state
\begin{equation} \label{eos}
P = K \rho^{1 + 1/n},
\end{equation}
where $P$ is the pressure, $\rho$ the rest mass density, $n$ the
polytropic index and $K$ the polytropic constant, $\kappa_n$ can be
derived from the Lane-Emden function (see, e.g,
LRS). Equation~(\ref{dotJ1}) can then be written
\begin{equation} \label{dotJ2}
\dot J = - f(e) \frac{5}{3 \kappa_n} \,\frac{J}{M}\, L.
\end{equation}

From equation~(\ref{dotJ2}), we find that the angular momentum $J$
changes by at most (Huet, 1996)
\begin{equation} \label{deltaJ}
\frac{\Delta J}{J} \sim - \frac{5}{3 \kappa_n}\, \frac{U_{\rm hot}}{M},
\end{equation}
assuming that the merged remnant radiates away all of its thermal
energy $U_{\rm hot} \sim L \Delta \tau$ prior to collapse.  The
ratio $U_{\rm hot}/M$ following merger can be estimated by taking the
progenitor stars to be cold, with $U_{\rm hot} = 0$, prior to reaching
the ISCO. During the subsequent plunge and merger, a part of the
kinetic energy will be converted into thermal energy by contact shocks
(Rasio \& Shapiro, 1994; Ruffert, Janka \& Sch\"afer, 1996). In the
extreme case of a head-on collision, a very large part of the kinetic
energy will be converted into heat. The ISCO, however, usually occurs
at a very small separation of the stars, and therefore we expect that
only a small part of the kinetic energy will be converted into heat.

To estimate $U_{\rm hot}$, we construct post-Newtonian, compressible
ellipsoid models of hot, uniformly rotating stars, generalizing the
cold post-Newtonian models of Lombardi, Rasio and Shapiro (1997).
Matching these configurations to binary models at the ISCO calculated
by Baumgarte {\em et.~al.} (1997b), we typically find $U_{\rm
hot}/U_{\rm cold} \sim 0.2 - 0.3$, where $U_{\rm cold}$ is the total
internal energy, excluding the thermal energy (i.~e.~kinetic energy of
degenerate nucleons; see Section~3 and Appendix A).  Since for typical
neutron stars $U_{\rm cold}/M \sim 0.1$, we have $U_{\rm hot}/M \sim$
few percent following merger.  Hence, from equation~(\ref{deltaJ}), we
expect that the relative change of the angular momentum due to
neutrino emission can be at most a few percent. We conclude that
neutrinos are very inefficient in reducing the merged remnant's
angular momentum.

We now focus on the change in $J/M^2$ associated with the neutrino
luminosity $L$:
\bea \label{cri0}
\frac{M^2}{J} \frac{d}{d\tau} \frac{J}{M^2} & = &
	\frac{\dot J}{J} - 2 \frac{\dot M}{M} \nn
	& = & \frac{\dot J}{J} \left(1 - 4\, \frac{T_{\rm rot}}{M} - 
		\frac{6}{5}\,\frac{\kappa_n}{u^t f(e)} \right).
\eea
Here we have used equations~(\ref{fl}) and~(\ref{dotJ2}) as well as
$T_{\rm rot} = \Omega J/2$. The magnitude of $J/M^2$ therefore {\em
increases} if
\begin{equation} \label{cri1}
1 - 4 \,\frac{T_{\rm rot}}{M} - \frac{6}{5}\,\frac{\kappa_n}{u^t f(e)} < 0.
\end{equation}
For a dynamically stable rotating ellipsoid with rotational
velocity $v \ll 1$ and $T_{\rm rot}/|W| \lesssim 0.27$ (see below),
we have $T_{\rm rot}/M \ll 1$, $u^t \sim 1$ and $f(e) \sim 1$, so that 
this criterion reduces to
\begin{equation}
\kappa_n > \frac{5}{6}.
\end{equation}
The critical value of $\kappa_n = 5/6$ corresponds to a polytropic
index of $n_{\rm crit} = 0.45$ (see, e.g., LRS, Table 1). Neutrino
radiation thus leads to an {\em increase} of the angular momentum
parameter $J/M^2$ for all equations of state with $n < n_{\rm
crit}$. Interestingly, the value of $n_{\rm crit} = 0.45$ is very
close to typical values that are expected in realistic neutron star
equations of state. Note that we have derived this numerical value in
the limit of slow rotation. For rapid rotation, as in the numerical
example of Section 3, we find that the value for $n_{\rm crit}$
increases in the ellipsoidal approximation.

For values of $n$ close to $n_{\rm crit}$, the terms on the left
hand side of equation~(\ref{cri1}) will always be close to zero.
According to equation~(\ref{cri0}), the fractional
change of $J/M^2$ will therefore be much smaller than the fractional
change of $J$, which we have estimated to be in the order of a few
percent at most. Nevertheless, it is interesting to note that the
emission of neutrinos can lead to an increase in $J/M^2$, and that for
nearly Newtonian configurations the sign of the change is
determined essentially by the stiffness of the equation of state.

\section{A NUMERICAL EXAMPLE}

\begin{figure}[t]
\epsfxsize=2.5in
\begin{center}
\leavevmode
\epsffile{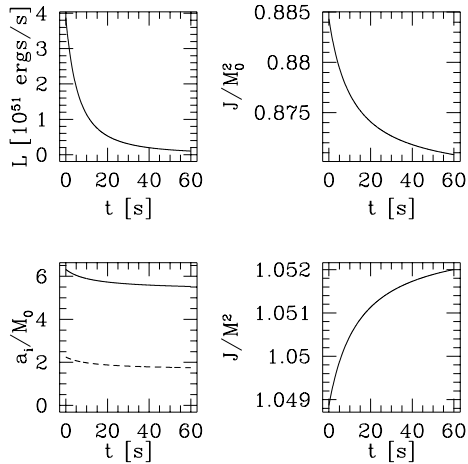}
\end{center}
\caption{Time evolution of a rapidly rotating neutron star in a
post-Newtonian ellipsoidal approximation (Model~1, see text).  We show
the luminosity $L$, the angular momentum $J/M_0^2$, the principle axes
$a_1$ (solid line in the lower left panel) and $a_3$ (dashed line),
and the angular momentum parameter $J/M^2$. Although the merged
remnant in Model~1 exceeds the maximum allowed mass of a cold, 
nonrotating configurations, it is stabilized against collapse by 
angular momentum.}
\end{figure}

To illustrate these effects, we dynamically evolve axisymmetric
post-Newtonian, compressible ellipsoid models of hot neutron star
merger remnants. We employ the formalism of Lai, Rasio \& Shapiro
(1994), supplemented by post-Newtonian corrections (Lombardi, Rasio \&
Shapiro, 1997) and thermal contributions to the internal energy. The
thermal corrections are taken into account by naively decomposing the
polytropic constant $K$ in~(\ref{eos}) into a linear sum of a cold and
a hot contribution, $K = K_{\rm cold} + K_{\rm hot}$.  The complete
formalism and equations can be found in Appendix~A.

\begin{figure}[t]
\epsfxsize=2.5in
\begin{center}
\leavevmode
\epsffile{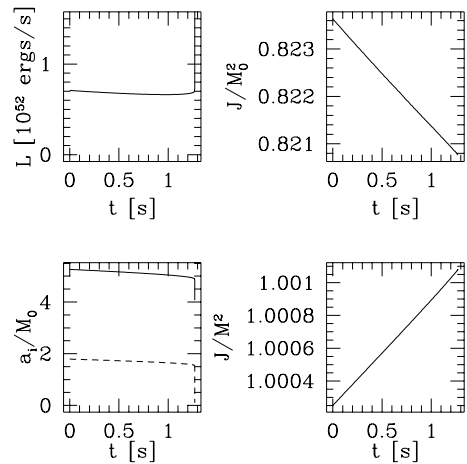}
\end{center}
\caption{Evolution of Model 2 (see text). The plots are labeled as in 
Figure~1.  Although Model~2 is initially stabilized against collapse by a
combination of thermal pressure and angular momentum, a delayed
collapse is induced as soon as it radiates away a small part of its
thermal energy.  Note that the neutrino luminosity $L$ {\em increases}
during the collapse.}
\end{figure}

For initial data we construct equilibrium configurations as described
in Appendix~A.3. These models of hot, rotating neutron stars form a
three parameter family, which can be uniquely determined by the
central density, the eccentricity and the thermal heat content
$(K_{\rm hot}/K_{\rm cold})$. We can choose these three parameters
such that the stars' rest mass $M_0$, total energy $M$ and angular
momentum $J$ match those of binary configurations at the ISCO as
calculated by Baumgarte {\em et.~al} (1997a,b). Here, we summarize
results for two different progenitor binary models with $n = 1$ and
$J/M^2 \geq 1$ (see lines 5 and 6 of Table 2 in Baumgarte {\em
et.~al.}, 1997b, for the binary parameters).  The two models differ by
how close the stellar masses are to the maximum allowed mass of an
isolated star in spherical equilibrium. We pick these two particular
cases since their evolution turns out to be qualitatively different.
For both models, we fix $K_{\rm cold}$ such that the rest mass of the
remant is $M_0 = 2 \times 1.5 M_{\odot}$, since observed binary
neutron stars all have gravitational masses close to 1.4 $M_{\odot}$,
corresponding to rest masses $\sim 1.5 M_{\odot}$. $K_{\rm cold}$ is
determined by the relation $K_{\rm cold} = (M_0/\bar M_0)^{2/n}$,
where the nondimensional quantity $\bar M_0$ is given in Table 2.

For Model 1 (line 5 of Table 2 in Baumgarte {\em et.~al.}, 1997b), we
have $K_{\rm cold} = 208 \mbox{~km}^2$.  In our post-Newtonian
approximation, this yields a maximum allowed rest mass of $M_0^{\rm
max} = 1.92 M_{\odot}$ (corresponding to a maximum allowed
gravitational mass of $M^{\rm max} = 1.71 M_{\odot}$) for isolated,
nonrotating cold stars.  These values are consistent with those
obtained from recent, realistic nuclear equations of state (see, e.g.,
Pandharipande, 1997). The initial gravitational energy of Model 1 is
$M = 2.75 M_{\odot}$ and the initial angular momentum parameter is
$J/M^2 = 1.05$. The remnant has an eccentricity of $e = 0.935$, a
thermal heat content $K_{\rm hot}/K_{\rm cold} = 0.25$ and $a_1 =
28.0$ km. The thermal energy in this model corresponds to a maximum
temperature of $\sim 14$ MeV.

For Model 2 (line 6), we have $K_{\rm cold} = 173 \mbox{~km}^2$,
corresponding to $M_0^{\rm max} = 1.75 M_{\odot}$ and $M^{\rm max} =
1.56 M_{\odot}$, so that this model's rest mass exceeds the maximum
allowed rest mass by a larger amount than Model 1.  It has an initial
gravitational mass of $M = 2.72 M_{\odot}$ and $J/M^2 = 1.0$,
corresponding to a remnant of $e = 0.94$, $K_{\rm hot}/K_{\rm cold} =
0.33$, and $a_1 = 23.3$ km. Here the maximum temperature is $\sim 28$
MeV, which is similar to typical values found by Ruffert, Janka \&
Sch\"afer (1996).

We show the dynamical evolution of these two models in Figures~1
and~2.  Even though the rest mass of Model~1 greatly exceeds $M_0^{\rm
max}$, it is dynamically stabilized, within our approximations, by its
high angular momentum, even after all the thermal energy has been
radiated away. By contrast, Model~2 collapses after it
has emitted part of its thermal energy. In Model~1, the angular
momentum decreases by $\sim 1.5\%$, while $J/M^2$ {\em increases} by
$\sim 0.3 \%$. This shows that post-Newtonian corrections in an
ellipsoid approximation, as well as rotation and deviations from
sphericity, cause the the critical polytropic index $n_{\rm crit}$ to
increase to a value greater than one. In Model~2, the changes in $J$
and $J/M^2$ show the same trends but are smaller, because the star
emits only a small part of the thermal energy before it collapses.
In a fully relativistic treatment, the collapse would not proceed
directly to a black hole, because the Kerr limit on $J/M^2$ would
be exceeded. Binaries with masses closer to the maximum nonrotating
mass have $J/M^2 < 1$ (see Table 2 in in Baumgarte {\em et.~al.}, 1997b). 
Their remnants could collapse directly to Kerr black holes.

Note also that the neutrino luminosity {\em increases} as Model~2
undergoes 'delayed' collapse. This is in contrast to
spherically symmetric results for delayed collapse (Baumgarte, Shapiro
\& Teukolsky, 1996; Baumgarte {\em et.~al.}, 1996), in which the
luminosity always {\em decreases} during collapse. This difference can
be understood from a simple geometrical scaling argument.  As
explained in detail in Appendix~A, we assume for simplicity that we 
can write the thermal energy in the polytropic form
$\epsilon_{\rm hot} = n K_{\rm hot}
\rho^{1 + 1/n}$ (equation~(\ref{eps_therm})). In the high temperature 
limit, $\epsilon_{\rm hot} \sim \epsilon_{\rm rad}$, so that $T^4 \sim
\rho^{1 + 1/n} \sim (a_1^2a_3)^{-(1 + 1/n)}$.  Taking the surface area
of the ellipsoid to be $\sim 4 \pi a_1^2$ we find from~(\ref{L}) that in
the diffusion approximation the luminosity scales as
\begin{equation}
L \sim a_1^{3 - 1/n} a_3^{-(1 + 1/n)/2}
\end{equation}
For spherically symmetric collapse, $a_1 = a_3$ and $L \sim a_3^{(5 -
3/n)/2}$. For all $n > 3/5$ (which accomodates most realistic
equations of state), spherically symmetric collapse therefore leads to
a decrease of the luminosity: the neutrino optical thickness increases
faster than the thermal energy increases.  If, however, the star
collapses to a pancake, as in Figure~2, $a_1$ will remain finite, and
$L \sim a_3^{-(1 + 1/n)/2}$.  In this case the increase of the thermal
energy overcomes the increase of the optical thickness, and the
luminosity increases. Whenever the surface of a collapsing
star approaches a newly formed event horizon, the luminosity will
always be suppressed by the increasing redshift and black hole
capture. These effects are absent in our model calculations.
Nevertheless, our results suggest that the collapse of a rapidly
rotating, hot neutron star following merger may leave a very distinct
signature in the neutrino signal.

\section{ESTIMATE OF TIMESCALES}

So far we have focussed on the emission of neutrinos and their role in
the evolution of the merged remnant. To estimate if this evolution is
indeed governed by neutrinos, we now compare the timescale for the
neutrino emission with those associated with eletromagnetic and
gravitational radiation.

The timescale for the emission of the neutrinos is given by the
diffusion time, $\tau_{\nu}$. Taking $\kappa$ to be Rosseland mean
opacity due to scattering off nondegenerate neutrons and protons (see,
e.g., Baumgarte, Shapiro \& Teukolsky, 1996), we find
\bea
\tau_{\nu} & \sim & \kappa \rho R^2 \sim
	\kappa \frac{M}{R} \\[1mm]
	& \sim & 4 \left(\frac{kT}{20\,\mbox{MeV}} \right)^2
	\left(\frac{M}{2 M_{\odot}} \right)
	\left(\frac{R}{15 \mbox{km}} \right)^{-1} \mbox{s}.
	\nonumber
\eea

Assuming that electromagnetic emission is dominated by a magnetic
dipole radiation, its timescale can be estimated from (see, e.g.,
Shapiro \& Teukolsky, 1983)
\bea
\tau_{\rm EM} & \sim & \frac{T_{\rm rot}}{\dot E} \sim 
	\frac{M}{3 B^2 \sin^2 \alpha R^4 \Omega^2} \nn
	& \sim &
	7 \times 10^7 \left( \frac{M}{2 M_{\odot}} \right)	
	\left( \frac{B \sin \alpha}{10^{12} \mbox{G}} \right)^{-2} \nn
	& & \left( \frac{R}{15 \mbox{km}} \right)^{-4} 
	\left( \frac{\Omega}{10^4 \mbox{s}^{-1}} \right)^{-2} \mbox{s}.
\eea
Clearly, $\tau_{\rm EM}$ is by many orders of magnitudes larger than
$\tau_{\nu}$. Electromagnetic radiation is therefore negligible in the
early evolution of hot neutron stars, unless the magnetic fields are
extreme ($B \gtrsim 10^{15}$ G).

An axisymmetric, stationary star does not emit gravitational
waves. However, if the ratio of the kinetic and potential energy $t
\equiv T_{\rm rot}/|W|$ is large enough, the star will develop a bar
instability and will then emit gravitational radiation. More
specifically, the star is secularly unstable to the formation of a bar
mode if $t > t_{\rm sec} \sim 0.14$, and dynamically unstable if $t >
t_{\rm dyn}
\sim 0.27$ (Chandrasekhar, 1969; Shapiro \& Teukolsky, 1983; see
Stergioulas \& Friedman, 1997, and Friedman \& Morsink, 1997, for a
recent relativistic treatment). Writing $t$ in terms of the
ellipsoidal expressions~(\ref{w}) and~(\ref{t_rot}), we find
\bea \label{beta}
t & = & \frac{5(5-n)}{12 \kappa_n} \, 
	\frac{e(1-e)^{1/6}}{\sin^{-1} e} \,\left(\frac{J}{M^2}\right)^2
	\,\frac{M}{R} \nn
	& \sim & 2 \,\left(\frac{J}{M^2}\right)^2 \,\frac{M}{R}.
\eea
Since relativistic configurations typically have $M/R \gtrsim 0.1$,
this suggests that all neutron stars, rotating with $J/M^2 \gtrsim 1$,
have $t \gtrsim 0.2$ and are at least secularly unstable to the
formation of a bar mode. This is in agreement with the general
relativistic, numerical models of Cook, Shapiro \& Teukolsky
(1992). For $t_{\rm sec} < t < t_{\rm dyn}$, the timescale for the
formation of a bar mode driven by gravitational radiation can be
approximated by (see, e.g., Friedman \& Schutz, 1975; Lai \& Shapiro,
1995)
\begin{equation}
\tau_{\rm bar} \sim  3 \left( \frac{M}{2 M_{\odot}} \right)
	\left( \frac{M/R}{0.2} \right)^{-4} 
	\left( \frac{t - t_{\rm sec}}{0.1} \right)^{-5} \mbox{s}.
\end{equation}
Note that shear viscosity can also drive a bar mode instability, but
it is inefficient in hot neutron stars with $T \sim 10$ MeV as is the
case here (see Bonazzola, Frieben \& Gourgoulhon, 1996, and references
therein).  Once a bar has fully developed, the gravitational radiation
dissipation timescale can be estimated from the quadrupole emission
formula
\bea
&&\tau_{\rm grav} \sim \frac{T_{\rm rot}}{\dot E} \sim 
	\frac{I \Omega}{(M/R)^2 (R\Omega)^6} \sim 
	\frac{M}{(M/R)^2 (R\Omega)^4} \nn
	& &\sim
	4 \times 10^{-3} \left(\frac{M}{2M_{\odot}}\right) 
	\left( \frac{M/R}{0.2} \right)^{-2} 
	\left( \frac{R\Omega}{0.5} \right)^{-4} \mbox{s}.
\eea
For $t < t_{\rm dyn}$, the total gravitational radiation time\-scale
$\tau_{\rm GW} \sim \tau_{\rm bar} + \tau_{\rm grav}$ is therefore
dominated by the initial growth time $\tau_{\rm bar}$. Since
$\tau_{\rm bar}$ strongly depends on $t$, this timescale is quite
uncertain. However, the estimate~(\ref{beta}) suggests that for
compact stars with $J/M^2 \gtrsim 1$, we have $\tau_{\rm GW} \lesssim
3$ s.  We then see that the evolution of these merged stars is
characterized by the formation of a bar mode (see Lai \& Shapiro,
1995) and emission of gravitational waves, which may then carry off
angular momentum very efficiently. In some cases, $t > t_{\rm dyn}$,
and a bar may develop on a dynamical timescale ($\sim \Omega^{-1}
\sim$ few ms; see Rasio \& Shapiro (1994) for a numerical demonstration).

\section{CONCLUSIONS}

We conclude that neutrinos are very inefficient in carrying off
angular momentum from hot, massive remnants of neutron star binary
mergers, even if these are rapidly rotating and have an angular
momentum parameter $J/M^2 \gtrsim 1$.  Curiously, neutrino emission
may even {\em increase} $J/M^2$ by a small amount. We find that the
timescale for the formation of a bar mode may, for such objects, be
much shorter than the neutrino emission timescale. For pure
gravitational radiation the third term in equations~(\ref{cri0})
and~(\ref{cri1}) vanishes, so that $J/M^2$ always decreases.  This
suggests that gravitational waves will be faster and more efficient in
reducing $J/M^2$ than neutrinos. Together, these arguments may have
important consequences for gravitational wave detectors, since they
suggest that the evolution of mergered binary remnants will be
dominated by the emission of gravitational waves. In particular,
should $J/M^2$ be larger than unity upon merger, gravitational waves
seem to be the only means by which $J/M^2$ can be efficiently reduced
to a value smaller than unity, so that the entire merged configuration
can collapse to a black hole. Otherwise, hydrodynamic stresses will have
to support or expel some of the matter to allow the interior regions to
collapse.

These arguments imply that, except for cooling and deleptonizing,
neutrino emission plays a very minor role in determining the mass and
spin of the final configuration (black hole or rotating neutron
star). However, because of the role in inducing delayed collapse in
some cases, the neutrinos will have to be included in calculations of
the late stages of coalescence.  Gravitational radiation will play a
very important role in determining the final mass and spin of the
black hole or neutron star. This result strengthens the conclusion
that a fully general relativistic description of binary neutron star
mergers is essential for full understanding.

Note that our arguments apply to massive, hot, rotating neutron stars
in general, and not only to the remnants of binary neutron stars. For
example, they equally apply to newly formed, hot neutron stars in
supernovae. It has recently been suggested that these stars may be
stable initially, but may later deleptonize, undergo a phase
transition and collapse to a black hole (see, e.g., Brown \& Bethe,
1994).  Alternatively, such a delayed collapse might result from the
loss of angular momentum from a nascent, rapidly rotating neutron star
which is stabilized initially by its high spin. Neutrino emission
could, in principle, reduce this angular momentum and hence induce
collapse.  While for type II supernovae $U_{\rm hot}/M$ is slightly
larger than in the examples presented in Section~3, our arguments show
that the neutrino emission is still very inefficient in carrying off
angular momentum. This fact makes this scenario very unlikely (Huet,
1996). Similarly, our analysis suggests that neutrino emission plays a
negligible role in determining the quasi-steady spin rate of accreting
neutron stars at the center of Thorne-Zytkow objects, which have
recently been suggested to be likely sources of gravitational waves
for detection by GEO (Schutz, 1997).

\acknowledgments

It is a pleasure to thank Patrick Huet and Chris Pethick for
stimulating our interest in some of the issues discussed in this
paper, and Fred Lamb and James Lombardi for useful discussions. This
paper was supported in part by NSF Grant AST 96-18524 and NASA Grant
NAG 5-3420 to the University of Illinois at Urbana-Champaign.

\begin{appendix}

\section{POST-NEWTONIAN, ELLIPSOIDAL \break MODELS OF HOT, ROTATING NEUTRON
STARS}

\subsection{The Euler-Langrange Equations}

The total energy of an axisymmetric, rotating, hot neutron star can,
in a post-Newtonian, ellipsoidal approximation, be written as the
sum
\begin{equation} \label{energy}
M = U + U_{PN} + W + W_{PN} + T,
\end{equation}
where $U$ is the internal energy, $W$ the potential energy, $T$ the
kinetic energy, and $U_{PN}$ and $W_{PN}$ are post-Newtonian
corrections to the internal and potential energy. 

Including the post-Newtonian terms not only improves the solution
for relativistic stars quantitatively, but also, more importantly, it
changes the solution qualitatively. Without the post-Newtonian
corrections, a graph of the rest mass $M_0$ as a function of the
central density $\rho_c$ exhibits no turning
point. By including the post-Newtonian terms, we find turning points
reasonably close to the turning points of the corresponding fully
relativistic (Tolman-Oppenheimer-Volkov) solution (see Lombardi, Rasio
\& Shapiro, 1997).  Since these turning points give the maximum
allowed mass configuration, the post-Newtonian corrections allow us to
approximate the onset of radial instability and to mimic the delayed
collapse of hot, rotating neutron stars to black holes.

We assume a polytropic equation of state
\begin{equation}
P = K \rho^{1 + 1/n},
\end{equation}
and crudely account for thermal pressure by taking the constant $K$ to
have a cold contribution due to degenerate nucleons, $K_{\rm cold}$,
and a hot contribution due to thermal heating, $K_{\rm hot}$:
\begin{equation}
K = K_{\rm cold} + K_{\rm hot}.
\end{equation}
The internal thermal energy density $\epsilon_{\rm hot}$ is
therefore taken to be of the form
\begin{equation} \label{eps_therm}
\epsilon_{\rm hot} = n K_{\rm hot} \rho^{1 + 1/n}.
\end{equation}
While $K_{\rm cold}$ is assumed to be constant in both space and time,
we allow $K_{\rm hot}$ to depend on time, keeping it constant
throughout the star. Integrating the energy density over the star
allows us to write the total internal energy in terms of the central
density $\rho_c$ as a sum of a cold and a hot contribution
\begin{equation}
U = U_{\rm cold} + U_{\rm hot} 
	= k_1 (K_{\rm cold} + K_{\rm hot}) \rho_c^{1/n} M_0
\end{equation}
where $k_1$ is a dimensionless structure constant that depends on the
density profile, i.e.~the polytropic index $n$. Numerical values for
$k_1$ can be found, for example, in LRS (see Table~1).  In terms of
the mean radius
\begin{equation}
R \equiv (a_1^2 a_3)^{1/3},
\end{equation}
the central density can be written
\begin{equation}
\rho_c = \beta \frac{M_0}{R^3},
\end{equation}
where we have defined
\begin{equation}
\beta \equiv \frac{\xi_1}{4 \pi | \theta'(\xi_1) | },
\end{equation}
and where $\theta$ and $\xi$ are the usual Lane-Emden parameters for a 
polytrope (see, e.g., Chandrasekhar, 1939). The internal energy can
therefore be rewritten
\begin{equation}
U = k_1 K \beta^{1/n} \frac{M_0^{1 +1/n}}{R^{3/n}}.
\end{equation}

The potential energy $W$ is given by
\bea \label{w}
W & = & - \frac{3}{5-n} \frac{M_0^2}{R} \frac{\sin^{-1} e}{e} (1 - e^2)^{1/6}
	\\[1mm]
	& = & - \frac{3}{5-n} \beta^{-1/3} M_0^{5/3} \rho_c^{1/3} 
	\frac{\sin^{-1} e}{e} (1 - e^2)^{1/6}. \nonumber
\eea
Introducing the dimensionless coefficients $A_i$ (Chandrasekhar, 1969,
$\S 17$)
\begin{equation}
A_1 = (1 - e^2)^{1/2} \frac{\sin^{-1}e}{e^3} - \frac{1 - e^2}{e^2}
\end{equation}
and
\begin{equation}
A_3 = 2 - 2A_1
\end{equation}
(for $a_1 > a_3$), the potential energy can be rewritten
\begin{equation} 
W = - \frac{3}{5-n} \frac{M_0^2}{R} \frac{{\cal J}}{2 R^2},
\end{equation}
where
\begin{equation}
{\cal J} = 2 A_1 a_1^2 + A_3 a_3^2.
\end{equation}
The last term ${\cal J}/(2R^2)$ takes into account corrections to the
potential energy due to deviations from sphericity. Note that partial
derivatives of ${\cal J}$ take the form
\begin{equation}
\frac{\partial {\cal J}}{\partial a_i} = \frac{1}{a_i} ({\cal J} - a_i^2 A_i),
\end{equation}
so that
\begin{equation}
\frac{\partial W}{\partial a_i} = \frac{3}{5-n} \frac{M_0^2}{2 R^3} a_i A_i
\end{equation}
(see Chandrasekhar, 1969).

Assuming the star to be uniformly rotating, the kinetic energy $T$ can
written as a sum of a contribution due to expansion
\begin{equation}
T_{\rm exp} = \frac{1}{10} \kappa_n M_0 (2 \dot a_1^2 + \dot a_3^2)
\end{equation}
and a contribution due to spin
\begin{equation} \label{t_rot}
T_{\rm rot} = \frac{1}{2} I \Omega^2 = 
	\frac{5}{4 \kappa_n} \beta^{-2/3} J^2 M^{-5/3} \rho_c^{2/3} (1 -
	e^2)^{1/3}.
\end{equation}
Here the moment of inertia is given by
\begin{equation}
I = \frac{2}{5} \kappa_n M_0 a_1^2,
\end{equation}
and $\kappa_n$ is a structure constants of order unity that
depends only on the polytropic index $n$ (see, e.g., LRS)

Following Lombardi, Rasio \& Shapiro (1997), the post-Newtonian
corrections can be written
\bea
U_{PN} & = & - l_1 K \rho_c^{1/n + 1/3} M_0^{5/3} \nn
	& = & 
	-l_1 K \beta^{1/n + 1/3} \frac{M_0^{2 + 1/n}}{R^{1 + 3/n}}
\eea
and
\begin{equation}
W_{PN} = - l_2 \rho_c^{2/3} M_0^{7/3} = 
	- l_2 \beta^{2/3} \frac{M_0^3}{R^2}.
\end{equation}
Numerical values for the structure constants $l_1$ and $l_2$ are calculated
in Lombardi, Rasio \& Shapiro (1997). Both post-Newtonian terms are
correct in this form only for spherical configurations. However, we shall
assume the slow rotation limit ($T_{\rm rot}/|W| \ll 1$), for which 
corrections arising from deviations from sphericity are higher order
and will be neglected.

From these energy contributions we can now construct the Lagrangian
\begin{equation} \label{lagrangian}
L(q_i;\dot q_i) = T - U - U_{PN} - W - W_{PN},
\end{equation}
where the $q_i$ are the independent variables $a_i$ and $\phi$ 
(with $\dot \phi = \Omega$). The dynamics of the system is then
governed by the Euler-Lagrange equations
\begin{equation} \label{el}
\frac{d}{d\tau} \frac{\partial L}{\partial \dot q_i} 
	- \frac{\partial L}{\partial q_i} = 0.
\end{equation}
The angular momentum can be defined as
\begin{equation}
J \equiv \frac{\partial L}{\partial \Omega}.
\end{equation}
Since $\phi$ is an ignorable coordinate of the
La\-gran\-gi\-an~(\ref{lagrangian}), the Euler-Lagrange equations imply that
$J$ is conserved on a dynamical timescale. However, we want to allow
for the radiation of angular momentum on a secular timescale 
by the emission of neutrinos,
so we will therefore allow for an external dissipation term on the right
hand side of equation~(\ref{el}) when $q_i = \phi$:
\begin{equation}
\frac{dJ}{d\tau} = - (\dot J)_{\nu}.
\end{equation}
Note also that when we allow $K_{\rm hot}$ to decrease with time as the 
star cools (see equation~(\ref{dotKhot}) below), the system will no longer
conserve energy.

We now find that the Euler-Langrange equations yield
\begin{eqnarray} \label{dota1}
\ddot a_1 & = & \Omega^2 a_1 + K \beta^{1/n}
	\frac{5 k_1}{n \kappa_n} \frac{ M_0^{1/n}}{R^{3/n}a_1} - \\[1mm]
& &  	K \left(\frac{1}{n} + \frac{1}{3}\right) \beta^{1/n + 1/3} 
	\frac{5 l_1}{\kappa_n} \frac{M_0^{1/n+1}}{R^{3/n+1}a_1} - \nn
& & 	\frac{10}{3} \beta^{2/3} \frac{l_2}{\kappa_n} \frac{M_0^2}{R^2 a_1}
 	- \frac{5}{2\kappa_n} \frac{3}{5 - n}\frac{M_0}{R^3}a_1 A_1, \nn
\ddot a_3 & = & K \beta^{1/n}
	\frac{5 k_1}{n \kappa_n} \frac{ M_0^{1/n}}{R^{3/n}} - \\[1mm]
& & 	K \left(\frac{1}{n} + \frac{1}{3}\right) \beta^{1/n + 1/3} 
	\frac{5 l_1}{\kappa_n} \frac{M_0^{1/n+1}}{R^{3/n+1}a_3} - \nn
& &	\frac{10}{3} \beta^{2/3} \frac{l_2}{\kappa_n} \frac{M_0^2}{R^2 a_3}
	- \frac{5}{2\kappa_n} \frac{3}{5 - n}\frac{M_0}{R^3}a_3 A_3, \nonumber
\end{eqnarray}
and
\begin{equation} \label{dotOmega}
\dot \Omega = -\frac{1}{I}(\dot J)_{\nu} - 2\frac{a_1 \dot a_1}{a_1^2}\Omega.
\end{equation}
In the absence of neutrino emission ($(\dot J)_{\nu} = 0$), these
ordinary differential equations completely describe the adiabatic
evolution of a uniformly rotating, axisymmetric neutron star in a
post-Newtonian ellipsoid approximation. They represent in this limit
the ellipsoidal analogue of the post-Newtonian equations of
hydrodynamics for an adiabadic gas.  In order to allow for cooling and
loss of angular momentum, we have to supplement this system of
equations with expressions for $\dot K_{\rm hot}$ and $(\dot
J)_{\nu}$.

\subsection{Neutrino Cooling and Angular Momentum Loss}

We approximate the thermal energy contribution from hot nucleons by
\begin{equation}
\epsilon_{\rm nucl} = \frac{3}{2}\,nkT
\end{equation}
(strictly true in the limit of high temperatures), and add to it the
thermal radiation of photons, electron-positron pairs and $N_{\nu} =
3$ flavors of (nondegenerate) neutrinos
\begin{equation}
\epsilon_{\rm rad} = (3 + 7N_{\nu}/8)aT^4.
\end{equation}
The total thermal energy density is then given by
\begin{equation}
\epsilon_{\rm hot} = \epsilon_{\rm rad} + \epsilon_{\rm nucl}.
\end{equation}
This expression can be used to estimate the temperature $T$.  Adopting
the Rosseland mean opacity arising from scattering off nondegenerate
neutrons and protons, the neutrino opacity takes the form
\begin{equation}
\kappa = \frac{7(2\pi)^2}{20} \frac{\sigma_0}{4 m_{\rm B}}
	\left( \frac{kT}{m_e} \right)^2,
\end{equation}
where $m_{\rm B}$ is the baryon mass, $m_e$ the electron
mass, and $\sigma_0 = 1.76 \times 10^{-44} \mbox{cm}^2$. The neutrino
luminosity can then be estimated via the diffusion formula
\begin{equation} \label{L}
L = \frac{{\cal A}}{3 \rho_a \kappa} |\nabla \epsilon_{\rm rad}|
	\sim \frac{{\cal A}}{3 \rho_a \kappa} \,
        \frac{\epsilon_{\rm rad}}{a_3},
\end{equation}
where $\rho_a = 3M_0/(4 \pi R^3)$ is the average density.  Since $a_3
< a_1$, we expect that the ``preferred'' direction of diffusion is
along the rotation axis, and therefore estimate the gradient of
$\epsilon_{\rm rad}$ by dividing $\epsilon_{\rm rad}$ by $a_3$.  Using
equation~(\ref{dotJ1}), the angular momentum loss can now be written
\begin{equation} \label{dotJ3}
(\dot J)_{\nu} = \,<\varpi^2> \Omega L.
\end{equation}

Next we have to relate the luminosity $L$ to $K_{\rm hot}$. From
equation~(\ref{defL}) we have
\begin{equation} \label{dotKhot0}
L = \frac{\delta W^{(e)}}{d \tau} = - <T\frac{dS}{d\tau}>,
\end{equation}
where the brackets denote an average over the star and $S$ is the
total entropy of the star. From the first law of thermodynamics we
have
\begin{equation}
Tds = d \left( \frac{\epsilon}{\rho}\right) + 
	P d \left( \frac{1}{\rho}\right).
\end{equation}
Here $\epsilon = n K \rho^{1 + 1/n}$ is the total internal energy
density, and $s$ is the specific entropy (per rest mass). For
adiabatic changes, $K$ is constant and $Tds = 0$.  Here, $K = K_{\rm
cold} + K_{\rm hot}$ is not constant, however, and we find
\begin{equation}
Tds = n \rho^{1/n} dK.
\end{equation}
In a Newtonian treatment, integration over the whole star now yields
\bea \label{tds}
<TdS>\, & = & \int (Tds) dm = dK \int n \rho^{1/n} dm \nn 
	& = & d \ln K \int \epsilon dm
	= U d \ln K.
\eea
In keeping with our earlier assumption, we take $dK$ to be constrant
throughout the star. To incorporate the post-Newtonian correction
consistently, we simply replace $U$ with $U + U_{PN}$.  Inserting this
result into~(\ref{dotKhot0}) then yields
\begin{equation} \label{dotKhot}
\dot K_{\rm hot} = - \frac{K}{U+U_{PN}} L.
\end{equation}
Equations~(\ref{dota1}) to~(\ref{dotOmega}) together with~(\ref{dotJ3}) 
and (\ref{dotKhot}) now approximate the evolution of a cooling, rotating
neutron star. At each timestep, the luminosity $L$ can be estimated
from~(\ref{L}).

Alternatively, we could have derived equation~(\ref{tds}) from
observing that, for quasi-static changes of constant rest mass, the
mass-energy of an equilibrium configuration changes according to
\begin{equation} \label{dm1}
dM = \Omega dJ + \Lambda d {\cal C} + <TdS>,
\end{equation}
where we have set $u^t \sim 1$, and where for completeness we have
allowed a vorticity term $\Lambda d {\cal C}$.  Here ${\cal C}$ is the
circulation and $\Lambda$ is the vorticity of the configuration.
Following Appendix~D in LRS, this change can also be written
\begin{equation}
d M = \left. \frac{\partial M}{\partial J} \right|_{\alpha_i, {\cal C}} dJ 
+ \left. \frac{\partial M}{\partial {\cal C}} \right|_{\alpha_i, J}d{\cal C}
+ \left. \frac{\partial M}{\partial \alpha_i} \right|_{\alpha_j \ne \alpha_i,
	J, {\cal C}} d\alpha_i.
\end{equation}
Here the $\alpha_i$ are the independent parameters of the configuration,
which we can take to be $\rho_c$, $e$ and $K_{\rm hot}$. 
Note that
\begin{equation}
\left. \frac{\partial M}{\partial J} \right|_{\alpha_i, {\cal C}} = \Omega
\end{equation}
and
\begin{equation}
\left. \frac{\partial M}{\partial {\cal C}} \right|_{\alpha_i, J} = \Lambda
\end{equation}
(see LRS).  Furthermore, since we are considering quasi-static changes
between equilibrium configurations, first-order variations with
respect to the dynamical parameters vanish, so
\begin{equation}
\left. \frac{\partial M}{\partial \rho_c} \right|_{e,K_{\rm hot},J,{\cal C}} 
= \left. \frac{\partial M}{\partial e} \right|_{\rho_c,K_{\rm hot},J,{\cal C}}
= 0
\end{equation}
(see Section~A.3). Finally, from~(\ref{energy}),
\begin{equation}
\left. \frac{\partial M}{\partial K_{\rm hot}} \right|_{\rho_c,e,J,{\cal C}} 
= \frac{U + U_{PN}}{K}.
\end{equation}
Putting terms together, we therefore find 
\begin{equation}
dM = \Omega dJ + \Lambda d {\cal C} + (U + U_{PN}) d\ln K.
\end{equation}
Comparing this expression with~(\ref{dm1}) shows that we can identify
\begin{equation}
<TdS>\, = (U + U_{PN}) d \ln K,
\end{equation}
which agrees with~(\ref{tds}) (with $U$ replaced by $U + U_{PN}$).

\subsection{Equilibrium Configurations}

A configuration in equilibrium has $T_{\rm exp} = 0$, and the
equilibrium relations can then be found, for example, by setting the
derivatives of the energy functional~(\ref{energy}) with respect to
the central density $\rho_c$ and the eccentricity $e$ to zero:
\begin{equation}
\frac{\partial M}{\partial \rho_c} = \frac{\partial M}{\partial e} = 0.
\end{equation}
The first condition yields the virial relation
\begin{equation} \label{cond1}
\frac{1}{n} U + \frac{3 + n}{3n} U_{PN} + \frac{1}{3} W + \frac{2}{3} W_{PN}
	+ \frac{2}{3} T_{\rm rot} = 0.
\end{equation}
Since only $W$ and $T_{\rm rot}$ depend on the eccentricity $e$, the
second condition yields in this approximation the same result as a
purely Newtonian treatment (see equation~(3.21) in LRS)
\begin{equation} \label{cond2}
t = \frac{T}{|W|} = \frac{3}{2e^2} \left( 1 - 
	\frac{e\,(1-e^2)^{1/2}}{\sin^{-1} e} \right) - 1.
\end{equation}
Note that $t$ only depends on the eccentricity, and is independent of,
for example, the mass or the polytropic index $n$.

For a given central density $\rho_c$, eccentricity $e$ and thermal
contribution $K_{\rm hot}$, equilibrium conditions can now be constructed
by first finding $t$ from equation~(\ref{cond2}). This value can 
then be inserted into~(\ref{cond1}) to eliminate $T_{\rm rot}$:
\begin{equation}
\frac{1}{n} U + \frac{3 + n}{3n} U_{PN} + \frac{1}{3} W (1 - 2t) 
+ \frac{2}{3} W_{PN} = 0.
\end{equation}
Writing the individual energies in terms of the rest mass $M_0$ and
the central density $\rho_c$, we obtain a quadratic equation for
$M_0^{2/3}$
\begin{equation}
A M_0^{4/3} + B M_0^{2/3} - C = 0
\end{equation}
or
\begin{equation} \label{M_0}
M_0^{2/3} = - \frac{B}{2A} \pm \left(\frac{B^2}{4A^2} + \frac{C}{A}
	\right)^{1/2},
\end{equation}
where
\begin{eqnarray}
A & = & \displaystyle \frac{2}{3} l_2 \rho_c^{-1/3} \nn 
B & = & \displaystyle \frac{1}{5 - n} \beta^{-1/3} \rho_c^{-2/3}
	\frac{\sin^{-1}e}{e} (1 - e^2)^{1/6} (1 - 2t) + \nn
& &	 \frac{3 + n}{3n} l_1 K \rho_c^{1/n - 2/3} \\[1mm]
C & = & \displaystyle \frac{1}{n} k_1 K \rho_c^{1/n - 1} \nonumber.
\end{eqnarray}
In~(\ref{M_0}), the smallest positive solution is the physically
relevant solution. In the Newtonian limit, $A = 0$, and $M_0$ is given
by $(C/B)^{3/2}$ (see equation~(3.22) in LRS).  Once $M_0$ has been
determined as a function of $\rho_c$, $e$ and $K_{\rm hot}$, the
individual energy terms as well as the total energy can be calculated.

\end{appendix}

\end{document}